# Sudden acceleration of flames in open channels driven by hydraulic resistance

J. Yanez[1]*, M. Kuznetsov[1], V. Bykov[1]

[1] Karlsruhe Institute of Technology, 76131 Karlsruhe, Germany

*Corresponding author: jorge.yanez@kit.edu

Hydrogen-air deflagrations with venting at the end of obstructed tubes are studied experimentally and numerically. A *shockless* transition to the so-called *chocked regime* of the flame propagation is reported. Mixtures with 13% vol. of hydrogen were ignited from the open end of the tube at the interface between fuel and the ambient air. Three venting ratios were selected, closed, 40% and 100%. In all cases the flame initially propagates without acceleration at a velocity close to the laminar flame speed. The flame configuration excludes most of conventionally acknowledged phenomena of the DDT, namely, volumetric explosions, igniting shock and shock waves interactions. However, after an induction period, of the order of 1 sec, the flame accelerates more than 100 times, within a period of 3-30 ms, until the steady-state *choked* regime is established. The mechanism of such rapid acceleration is investigated both numerically and analytically. A one dimensional reduced description was suggested and analyzed to model the process of the flame acceleration. The study of the over simplified model reveals that the hydrodynamic resistance of the tube causes sudden flame acceleration and governs initial stage of the DDT.

## 1. Introduction

Deflagration experiments in tubes are usually initiated by a weak source of energy, like a spark, which produces the ignition of the reactive mixture in one of the extremes of the channel. The flame propagates slowly in the beginning with a velocity that depends on the mixture reactivity and may vary between several centimeters and several meters per second. The presence of obstacles in a confined media affects the deflagration. The local expansion of the burned gases in close proximity to obstacles generates turbulence that in a feedback mechanism increases the effective burning rate causing an acceleration of the propagation of the flame. This cycle drives the flame acceleration leading to high pressures within the reaction zone and, under certain circumstances, to transition to detonation. This transition is very sensitive to small perturbations, for instance, for lower reactive mixtures this feedback loop may be countered, or even interrupted, by the tendency of the flame to quench due to stretch and heat losses. Furthermore, lateral or end venting would cause energy and momentum losses that may slow down the flame and prevent detonations and sonic propagation regimes.

Most of the previous investigations dedicated to the acceleration of flames in obstructed channels have devoted a special emphasis to the deflagration to detonation transition (DDT) e.g. Bradley et al. (2008) and Oran and Gamezo (2007). The problem of the transition from deflagration to detonation regimes of the flame propagation was investigated in closed tubes (half closed), where an ignition source was placed against the closed end.

Dorofeev et al. (2000 and 2001) and Kuznetsov et al. (2002) have developed some criteria to predict the flame propagation regime in tubular configurations. Following those authors, the expected regimes depend on the tube geometry, and specifically, on the configuration of the obstacles. The stationary modes of the confined flame propagation can be classified as *slow sub-sonic, sonic (choked), fast super-sonic* and *quasi-detonations* (due to the considerable momentum loss at the baffles the Chapman-Jouguet propagation velocity is not reached in obstructed tubes).

*Fast* and *quasi-detonations* regimes may be suppressed by the use of venting orifices. Ciccarelli et al. (1998) and Alekseev et al. (2001) experimentally investigated the effect of the lateral venting on flame acceleration and DDT. It was found that the amount of reactive mixture necessary for the development of sonic flames or DDT grows with an enlargement of the venting surface. A comparison between end- and side-vented explosions in tubes showed a higher efficiency of the end opening reducing the combustion pressure (Alexiou and Andrews 1997).

Although the most important stationary regimes (deflagration, fast subsonic, sonic and supersonic, chocking, quasi- and detonation regimes) their multiplicity and peculiarities have been studied extensively and understood quite well (see e.g. Brailovsky and Sivashinsky (2000) the transition between these combustion modes is still poorly understood and it represents very complex open problem.

In the current study flames propagating in obstructed channels ignited from its open end are investigated. The initial stage of the DDT – initial transition of the deflagration to fast subsonic flames is in the focus of the study. In such configuration, a prolonged *quasi-laminar* propagation phase is followed by a sudden and extremely violent *shockless* flame acceleration that culminates when the *sonic* regime is reached.

Actually, the main mechanism of this sudden acceleration of flames was suggested by Brailovsky and Sivashinsky (2000). However, in their studies the main focus was made on the transition of the fast sub/super-sonic flames to the detonation and on multiplicity of the detonation regimes themselves (quasi-detonation, fast flames driven by the diffusion of the pressure etc.). The conclusion obtained from their numerical analysis was that the ultimate cause triggering the DDT was the hydraulic resistance. In subsequent analytical and numerical studies (see e.g. Bykov et al. (2004), Sivashinsky (2007)) the role of the hydraulic resistance was further investigated and it was realized how markedly resistance affects the evolving flame.

The previous existing study dedicated to this non-typical flame acceleration behavior (see the simulations performed by Middha and Hansen (2008)), confirms the sudden transition, but it does not focus on driving critical mechanisms and does not include an analysis of the underlying physics.

It is clear that the existence of friction forces destroys the *quasi-laminar* deflagration wave as a stationary travelling wave. Nevertheless, the critical parameters driving this phenomenon and the form in which this destruction occurs remains unclear. In order to improve current understanding of the phenomenon, the authors have carried out the analysis presented in this article by combining experimental, numerical and analytical approaches to the problem.

**Figure 1.**

92  **2. Description of the experiments**

93  The experiments were performed in the DRIVER facility (Scholtyssek et al., 2000), which is an
94  obstructed combustion tube with a total length of 12.2 m and an internal diameter of 174 mm,
95  (see Figure 1). The degree of obstruction selected was equal to 0.6. A 13% vol. hydrogen-air
96  mixture at ambient conditions was ignited at the open end, directly at the interface between the
97  inflammable mixture and the surrounding air. The instrumentation included photo diodes and
98  pressure gauges installed along the channel. The venting ratio α of the orifice was varied from
99  0% (fully closed) up to 100% (fully open).

100 **Figure 2**

101 In Figure 2 the experimental distance-time (x-t) diagrams of the flame propagation for closed
102 (left) and vented channel (right) are shown, representing pressure and light records plotted
103 against time in vertical direction. In the closed channel (left), the flame accelerates immediately
104 after reaching the first obstacle generating an additional flow motion which steepens into the
105 shock wave. The turbulent flow ahead of the flame created by the thermal expansion of the
106 products supports the flame acceleration within a relatively short time after ignition (~0.1 s).
107 Beyond the run-up distance, of about 1.3 m, the flame reaches the *choked* regime and propagates
108 further with a steady velocity of 540 m/s (close to the sound velocity in the product) and with the
109 characteristic pressure of the leading shock wave oscillates from 6 to 11 bar.

110 In the presence of venting, combustion products are discharged in the atmosphere through the
111 end orifice (Fig. 2, right) and do not support flame acceleration. The flame propagates in a *quasi-*
112 *laminar* regime with a stationary velocity of 3.5 m/s. The experimental records show that during
113 this phase no significant pressure increments exist (≈300 Pa) and, therefore, no relevant flow
114 motion (generating turbulence) appears ahead of the flame. Nevertheless, around 1 s after the ignition, the flame suddenly accelerates from the *quasi-laminar* to the *choked* regime within the same interval of 3-30 ms. After the acceleration, the *choked* flame propagates until the end of tube with a constant velocity of 540 m/s creating peaks of pressure of 5-9 bar.

Experiments performed with different vent ratios showed that the run-up-distance, necessary to reach the sonic propagation, is proportional to the vent ratio. They established that even a fully open end cannot prevent flame acceleration to the sonic regime and demonstrate that the acceleration mechanism may trigger a DDT in vented tubes of larger diameters with more reactive mixtures.

## 3. Numerical simulation

Numerical simulations of the experiments were carried out with the combustion code *COM3D* (Kotchourko, et al. 1999) developed in the Karlsruhe Institute of Technology. In order to analyze the mechanism of the flame acceleration in presence of end discharge three values of the venting ratio were selected, 0%, 40% and 100%.

### 3.1. Modelling

The numerical representation of the problem include the geometry of the tube, the obstacles inside it and a supplementary volume with *open-non reflective* boundary conditions to simulate the release of the combustion products through the vent area into the atmosphere. The *open-non reflective* boundary conditions numerically reproduce the unconfined ambient air. The total volume and time to be simulated restrict the minimum resolution achievable to 5.8 mm due to the available computational power. Thus the Kolmogorov and Taylor turbulent micro-scales remain smaller than the grid size. Likewise, the integral scale of the turbulence is minor than the grid resolution. Moreover, the boundary layers remain largely unresolved making the use of wall

139   functions necessary. Furthermore, the laminar flame thickness is going to be lesser than the grid
140   size.
141   To overcome those restrictions, the KYLCOM combustion model (Yanez, 2010), specifically
142   designed for *under-resolved* calculations, and the standard k-ε turbulence model (Launder and
143   Sharma 1974) was utilized (restrictions of k-ε turbulence model appear i.e. Pope 2000). The
144   resolution adopted prevented the use of *Large Eddy Simulation* methods for the turbulence
145   modeling (for resolution requirements refer to (Jimenez 2004)). The initial levels of turbulence
146   and dissipation were chosen following the criteria of Arntzen (1998) making k and ε equal to
147   $1\cdot 10^{-4}$ $m^2/s^2$ and $m^2/s^3$ respectively. For the modeling of the combustion the *KYLCOM* model
148   (Yanez, 2010) was coupled with the turbulent burning velocity correlation proposed by Schmidt
149   et al. (1998).
150   The influence of the resolution in the flame and hydrodynamic instabilities requires further
151   considerations. Until the flame acceleration takes place, the *thermo-diffusive* instability plays an
152   important role (see Figure 3, left), as confirmed by the experimental data of Kuznetsov et al.
153   (1998) and the analysis Zeldovich et al. (1988).
154   **Figure 3**
155   Contrary to the case of closed tubes where wrinkling is meaningful only until the first obstacle is
156   reached, in tubes with venting, the *quasi-laminar* propagation region grows significantly and the
157   effect of flame folding becomes very important. Thus, to estimate the increase in burning
158   velocity due to this, the following relation due to Driscoll,

159   $$\Xi = \frac{1}{Le} \tag{1}$$

160  was utilized, where Ξ is the increase in burning velocity due to thermal-diffusive effects and Le
161  is the Lewis number of the mixture.
162  The *Kelvin-Helmholtz* instability is only partially resolved in the selected mesh. It may appear
163  (Kuznetsov, 1998) due to the interaction of the flame with obstacles (see Figure 3 right) but only
164  after the acceleration of the flame has started already. Therefore no special modeling was
165  considered for this effect.

167  **4. Results and analysis**
168      **4.1.    General discussion**
169  Figure 4 displays a comparison of the flame propagation obtained from the results of simulations
170  and experiments. The results obtained for the *closed* case shows the characteristic fast
171  acceleration typical of those problems. For the cases with venting, two regions with different
172  propagation regimes, *fast* and *slow*, can be clearly identified.
173  **Figure 4**
174  The repeatability of the essays was analyzed repeating the test with 40% of venting twice. The
175  results of the two experiments show significant divergence of the transition location confirming
176  high sensitivity of the transition to the initial conditions and systems parameters. However, the
177  order of magnitude of the transition time and location as well as all properties of the transition
178  (initial deflagration velocity, thickness of the sharp acceleration region etc.) remains akin. The
179  second test for 40% of venting and the experiment with 100% venting almost superpose and the
180  acceleration of the process suffers a 40% delay (from ~0.6 s to ~1.0 s after the ignition). The
181  results of the numerical simulation for 40% of venting appear in the interval between the two
182  experimental curves and therefore a positive agreement between both can be claimed.

183 In the test preformed with 100% of venting a significant delay in the transition to the *fast* flame
184 regime appear in the numerical calculations. The repeatability of the 100% venting case was not
185 analyzed experimentally, and therefore the time range uncertainty for the acceleration can be
186 estimated, by comparison with the 40% case, as the time registered in the experiment ±40%.
187 The time necessary for the change of combustion regime is another representative magnitude of
188 the problem. Its value was approximately 50 ms in the calculations, while in the experiments the
189 intervals between the gauges restrict the accuracy of the measurements to a range of 3-30 ms.
190 Although the authors admit the need in further detail studies to comprehend the discrepancies
191 between different experiments it is considered as one of the characteristic inherent to this type of
192 combustion problems.
193 In the *choked* propagation area (almost vertical sections), the results of the experiments and the
194 calculations agree rather well. The simulation of the closed case was already the object of a
195 previous investigation (Yanez et al. 2010). Therefore, further details about it are not included
196 here.
197
198    **4.2.    Borghi diagram analysis**
199 A preliminary insight into the occurring acceleration processes can be obtained with the help of
200 the Borghi diagram (Borghi, 1988). For the systematic mixture examination, the Karlovitz,
201 Damköhler, Lewis and Markstein numbers were calculated (see Table 1). The *Karlovitz* number,
202 was defined (Poinsot, 1991) as $Ka = u'/S_l \sqrt{v/u'L}$, the *Damköhler* number as 203    $Da = S_L^2 L / (\chi u')$, the *Lewis* number as $Le = \chi / D$, and the *Markstein* number (Zeldovich

204    1985), as $Ma = \dfrac{\chi - D}{\chi} \dfrac{E_a \left( T_b - T \right)}{2RT_b^2} + \dfrac{D}{\chi}$.

205 **Table 1.**

206 All magnitudes necessary to perform the analysis were obtained in the same location, slightly
207 ahead of the flame front. This location corresponds, for the *closed* and *vented* cases, to
208 acceleration and quasi-laminar propagation regime respectively. In the case of the *choked*
209 propagation, the data was sampled between the initial shock and the sonic flame.

210 **Figure 5**

211 Figure 5 shows two differentiated combustion regimes. In the *closed* case, a regular accelerating
212 flame is found (black diamond). The *quasi-laminar* propagations obtained for the cases with
213 venting (white symbols) are deeply inside the *laminar flame* region. During the flame
214 acceleration, the white dots will describe a vertical path, moving out of the *laminar* region,
215 traversing an area near the black diamond to finalize in the position of the *cross* symbol which
216 represents the *choked* regime. The location of the *cross*, inside the *thickened flames* region,
217 indicates that the flame is quite stable and confirms its capability to accelerate to the *sonic*
218 propagation regime. In a very short time, a flame propagation regime in which no interaction
219 with turbulence existed is substituted abruptly by a saturated turbulence interaction.

### 4.3. Analysis of the acceleration mechanism

222 While the flame penetrates in the tube, the combustion products are discharged into the
223 atmosphere via the venting orifice. The propagation of the flame inside the channel implies that
224 combustion products should traverse a longer distance until they are discharged suffering an

225 enhanced *momentum loss*. The hydrodynamic resistance may be expressed through the formula,
226 Brailovsky and Sivashinsky (2000)

227 $$F = -\frac{2c_D}{d}\rho u^2 \qquad (2)$$

228 in which $c_D$ is the drag force coefficient. The results of the numerical experiments carried out
229 with diverse flow velocities in the range 1-30 m/s allow approximating $c_D$ with the value 0.12.
230 Therefore the total *loss of momentum* can be estimated as

231 $$\Delta P = \int_0^{x_f} F dx$$

232 where integration is taken until the flame front position. The existence of obstacles increases the
233 complexity in the flow pattern. For the propagation of the flame in the laminar regime the
234 obstacles produce a change in the total surface of the flame and thus of the total fuel
235 consumption. In first approach this change can be estimated to be of the order of (1-BR) being
236 BR the Blockage Ratio. As the obstructions are gradually reached, cyclic oscillations in the
237 pressure (order of tens of Pa, peak of ~340 Pa, see Figure 6) as well as in the velocity of the
238 discharge products (see Figure 7) will appear. Those oscillations will have a frequency
239 $\omega_1 = \dot{x}_f / d$ where $\dot{x}_f$ is the velocity of the flame front and $d$ is the interval between obstacles,
240 which in this problem is equal to the diameter. Nevertheless, it is important to underline that no
241 shock waves develop during the entire quasi-laminar regime.

242 **Figure 6**

243 Figure 6 shows how pressure variations slightly compress and decompress the part of the tube
244 filled by the reactants. This area can be understood as a close cylinder, or a *drum*, in which the
245 flame actuates as an oscillating piston. In order to study the compression/decompression cycle of the reactants the one dimensional Euler's equations of continuity and impulse may be used to model the phenomenon. Performing cross derivatives on them (in t-x) and operating, the wave equation can be obtained supposing the velocity of the oscillations is small and thus the hydrodynamic resistance can be neglected. Additionally, by taking into account the observations performed during the numerical experiments, the oscillations inside the reactants resulted to be mainly of the first harmonic. The wave equation may thus be simplified to

$$\ddot{p} + \left(\frac{2\pi c}{4(L-x_f)}\right)^2 p = 0 \qquad (3)$$

where $L$ is the total size of the tube, c is the local sound velocity in the fresh mixture and $x_f$ is the position of the flame, and therefore a second cyclic process with a frequency

$$\omega_2 = \frac{c}{4(L-x_f)} \qquad (4)$$

is present in our system, as can be seen in Figure 6 and Figure 7.

**Figure 7**

The final pressure signal obtained, are the superposition of the two cyclic processes with frequencies $\omega_1$ and $\omega_2$, and the variable peaks of the registered amplitudes results from this superposition.

The resistance of the products grows linearly as the flame penetrates inside the tube (see Figure 6, thick line (trend)). When the resistance is comparable with the pressure peaks created by the flame, the products have difficulties to be discharged and a part of them are accumulated inside the tube. The reactants receive an enhanced compression and thus an increased compression-decompression cycle is triggered. The flame suffers an additional acceleration and traverses an augmented distance per oscillation. Some significant flow appears ahead of the flame. If during this displacement an obstacle is overcome, the burning rate will be enlarged by the turbulence created by the barrier and the flame starts to burn in the turbulent regime. The burning rate, the compression of the reactants and the hydrodynamic resistance are thus enhanced.

**Figure 8**

Next compression-decompression cycle (see Figure 8, left), will drive the flame to a very intense acceleration that will ultimately finish in the *choked* regime. Figure 8 (right) shows the behaviour of experimental light and pressure signals in the nearest proximity to the sonic flame transition point. At this position, the pressure oscillations due to the tube resistance become relatively strong initiating the mentioned mechanism.

Clearly, a smaller vent surface will reduce the run-up distance due to the enhanced *loss of momentum* on the orifice itself. Therefore, the critical value necessary to create overpressure and flow ahead of the flame will be achieved faster.

The coupling between the described phenomena is very complex. The small, but predictable, discrepancies between the repeated experiments with 40% of venting (Figure 4) caused by the distinct timing of the flame acceleration (i.e. the flame traverse the same length but only one obstacle is trespassed).

### 4.4. One dimensional reduced model

The discussion above hints on the possibility to use a one dimensional model of the propagation of the flame. In the following a coarse tube is considered to simplify the model and make it treatable analytically, in which the effect of the obstacles is taken into account as an enhanced hydrodynamic resistance. Two separate regions of the tube are considered for the modeling. In the so-called products region, between the flame and the discharge orifice the flow is assumed to

290  be uncompressible. For the reactants, region between the flame and the closed end of the tube,

291  the velocity is considered to be small and the term *u·(grad u)* can therefore be neglected during

292  the initial flame acceleration stage.

293  Thus, the equation of the momentum conservation

294  $$\frac{\partial \rho u}{\partial t} + \frac{\partial \rho u^2}{\partial x} + \frac{\partial p}{\partial x} = F. \tag{5}$$

295  for the region of the products, between the flame and the discharge orifice, becomes

296  $$\rho \frac{\partial u}{\partial t} + \frac{\partial p}{\partial x} = F. \tag{6}$$

297  Taking into account the open end and typical deflagration velocities before the flame

298  acceleration, uncompressible flow in the products region is assumed as well. In the case of

299  propagation of the flame in the deflagration laminar regime, and considering the reactants as

300  uncompressible, the velocity of the products can be defined as

301  $$u = -(\sigma - 1)\dot{x}_f, \tag{7}$$

302  with σ expansion ratio, which follows from the mass conservation and from the mean flame

303  surface velocity. Substituting (2) in (6) and integrating between the open end of the tube and the

304  position of the flame yields

305  $$-\rho \int_0^{x_f} (\sigma - 1)\ddot{x}_f dx + \int_0^{x_f} \frac{\partial p}{\partial x} dx = \int_0^{x_f} F dx \Rightarrow -\rho(\sigma - 1)\ddot{x}_f x_f + p_f - p_0 = \int_0^{x_f} F dx. \tag{8}$$

306  Substituting (2) in (8)

307  $$-\rho(\sigma - 1)x_f \ddot{x}_f + p_f - p_0 = -\frac{2c_D}{d}\rho(\sigma - 1)^2 \dot{x}_f^2 x_f \tag{9}$$

308  This equation contains $p_f$, pressure in the products side, as a free parameter that can be closed

309  with the equation (3) obtained for the pressure in the reactants area. The increment of pressure

310 between both sides, in the case of a stationary flame front can be calculated applying Rankine-
311 Hugoniot conditions

$$\Delta p \approx \rho\sigma(\sigma-1)\dot{x}_f^2 \tag{10}$$

**Figure 9**

314 With the help of this equation, the pressure in the reactants, $p_f^+$ side can be considered

$$\rho(\sigma-1)x_f\ddot{x}_f - p_f^+ + \rho\sigma(\sigma-1)\dot{x}_f^2 + p_0 = \frac{2c_D}{d}\rho(\sigma-1)^2\dot{x}_f^2 x_f \tag{11}$$

316 This equation can be re-written considering the over-pressure, $P$. If $\dot{x}_f(0) = S_L$ then the over-
317 pressure can be written as $P = p_f^+ - \rho\sigma(\sigma-1)S_L^2 - p_0$

$$\rho(\sigma-1)x_f\ddot{x}_f = P - \rho\sigma(\sigma-1)(\dot{x}_f^2 - S_L^2) + \frac{2c_D}{d}\rho(\sigma-1)^2\dot{x}_f^2 x_f \tag{12}$$

319 so that initially for $t=0$, $P(0)=0$.

320 This equation can be coupled with the equation (3) transformed for the over-pressure, which is

$$\ddot{P} + \left(\frac{2\pi c}{4(L-x_f)}\right)^2 P = 0, \tag{13}$$

322 to obtain a closed system. The result of this problem, considering as initial conditions $t=0$,
323 $x_f(0)=0.1$, when walls are reached by the flame with velocity $\dot{x}_f(0) = S_L$ are shown in Fig. 9. The
324 initial conditions for the pressure were obtained from the numerical experiments and where
325 $P(0)=0$ and $\dot{P}(0) = 10$. Although there is a good agreement of the results obtained with the one
326 dimensional simplification it has to be underlined that the validity of the analysis is restricted to
327 the initiation stage of the acceleration of the flame. Moreover, significant divergences obtained
328 in the reproduction of experiments themselves and of numeric simulations (compare critical

329  times shown in Figs. 4 and 9) may then be mathematically expressed through strong dependence

330  of the early flame development.

331  In order to illustrate this and explain the core mechanism of the flame acceleration let us consider

332  the equation (12) without the overpressure term, namely, $P(t)=0$ is assumed.

333  **Figure 10**

334  The system has the final form

$$\begin{cases} \ddot{x}_f x_f = -\sigma\left((\dot{x}_f)^2 - S_L^2\right) + \dfrac{2c_D(\sigma-1)}{d}(\dot{x}_f)^2 x_f \\ \dot{x}_f(0) = S_L \\ x_f(0) = x_f^0 = 0.1 \end{cases} \quad (14)$$

336  This system is studied in the phase plane by transforming the ODEs equation of second order to

337  a plane system of ODEs of the first order via regular transformation:

$$\begin{cases} v = x \\ u = \dot{x} \end{cases} \Rightarrow \begin{cases} \dot{v} = u \\ \dot{u}v = -\sigma(u - S^2_L) + \dfrac{2c_D(\sigma-1)}{d}u^2 v \end{cases} \quad (15)$$

339  The solution of the system in the phase plane coordinates looks

$$u = S_L\left(1 + \frac{4C_D(\sigma-1)}{d}\int_{x_f^0}^{v} s^{2\sigma} e^{-\frac{4C_D(\sigma-1)}{d}s} ds \, v^{-2\sigma} e^{\frac{4C_D(\sigma-1)}{d}v}\right)^{1/2} \quad (16)$$

341  Figure 10 shows the solution by the black dashed line. First observation concerns the limiting

342  behavior of the system solution in the vicinity of zero

$$\int_{x_f^0 \to 0}^{v} s^{2\sigma} ds \, v^{-2\sigma} \underset{v \to 0}{\approx} \frac{v}{2\sigma} \to 0 \quad (17)$$

344 demonstrates that $x_f^0 \to 0$ can be justified and no singularity occurs at the initial point. The second
345 and most important observation about the system initial behavior, namely, the role of the system
346 isocline of the flame speed equation:

$$\dot{u} = \left(-\sigma(u - S_L^2) + \frac{2c_D(\sigma-1)}{d}u^2 v\right)/v = 0 \Leftrightarrow u^2 = \frac{d\sigma}{2c_D(\sigma-1)}\left(\frac{S_L^2}{\frac{d\sigma}{2c_D(\sigma-1)} - v}\right) \quad (18)$$

348 Figure 10 shows by the solid black line that near the origin it represents a stable attractor, all
349 trajectories starting nearby converges (fast) to the lower branch of the isocline and follow the
350 detailed solution. Moreover, right after crossing the isocline the system solution trajectories
351 changes the character (speeding up instead of decreasing for initial point above the curve), this
352 make the border line which is asymptotically given by

$$v^* = x_f^* = \frac{d\sigma}{2c_D(\sigma-1)} = 0.94 \quad (19)$$

354 a very important and crucial property defining the critical behavior. It explains the transition
355 phenomena in terms of the phase plane. One clearly sees that if the initial point is on the right
356 from this curve $v = v^* = \frac{d\sigma}{2c_D(\sigma-1)}$ the vector field demonstrates the exponential increase of the
357 flame speed as a function of the flame distance.

359 Additionally, the form of the isocline dependence on the system parameters and variables

$$u^2 = \frac{d\sigma}{2c_D(\sigma-1)}\left(\frac{S_L^2}{\frac{d\sigma}{2c_D(\sigma-1)} - v}\right) \quad (20)$$

361 predicts the sensitivity of the critical phenomena on the initial pressure perturbation with respect
362 to time, but the sensitivity is lower with respect to the space. Indeed, Fig. 10 (right) shows that
363 there is no so much difference in the space if one replaces $S_L^2 \to \alpha S_L^2$ in the equation.
364 Qualitatively, similar phase portrait is observed, with the same critical value for the space (flame
365 position) with the main speed up of the flame in between $x_f^*$ and $2 x_f^*$, while numerical values
366 of the critical time equal $t_f^*(\alpha = 1) = 1.241897$ and $t_f^*(\alpha = 0.75) = 1.430550$ respectively. A
367 weak sensitivity to the perturbation of the initial pressure and the form of the critical parameter
368 can be explained in more simple way. Namely, the found asymptotic of the equation physically
369 means that the flame starts rapidly accelerating whenever the pressure jump (drop of the pressure
370 - work of the pressure force) less or equals to the work of the friction forces.

$$\Delta p = \rho \sigma (\sigma - 1) \dot{x}_f^2 \equiv \frac{2 c_D \rho_- (\sigma - 1)^2}{d} (\dot{x}_f)^2 x^*_f = \int_0^{x^*_f} F(u(t,s)) ds \to x^*_f = \frac{d \sigma}{2 c_D (\sigma - 1)} \quad (21)$$

372 Thus, when the work of the friction force starts dominate, the pressure in the reaction front
373 increases triggering the flame acceleration due to the cumulative effect of the pressure diffusion.

375 It is very important to note that there no regular singularity (reaching infinity in final time or
376 space as a reaction front position) was observed in the solution of the governing equations, just
377 very smooth and exponential (although hyper-geometric) growth of the system solution was
378 found to take place. This confirms, explains and fully justifies an irregular *shockless* character of
379 the flame acceleration observed in the experiments.

381 5. **Conclusions**

The problem of rapid acceleration of the flame in obstructed tubes with an open and vented end was analyzed experimentally, numerically and analytically. In order to study the acceleration problem of the flames propagating in tubes with different venting ratios the results of three experiments were simulated numerically. The dynamics of the combustion process was reproduced quite adequately by the numerical simulations. The obtained experimental results show that the deflagration propagation regime instantly accelerates, without generating shock waves, to the fast sonic flames inside a time interval of 3-30 ms. It was shown numerically that this sharp flame acceleration is a consequence of the coupling between three effects: the pressure oscillations of the closed space filled with reactants, the trespass of more than two obstacles by the flame in a single oscillation phase and the hydrodynamic resistance that depends on the total length passed by the products until they are discharged in the atmosphere. It was found that the latter has a pivotal role and to the leading order it defines the actual critical transition length, while the time of the transition is strongly influenced by the other factors. It has to be underlined that the observed acceleration process involves a pure hydrodynamic mechanism, which was confirmed by the analysis of one-dimensional simplified model. A parametric analysis performed using the Borghi diagram demonstrates the possibility of such scenario of the flame propagation. The same mechanism might be responsible for the detonation initiation in obstructed channels with end venting in case of more reactive mixtures or larger tube diameter.

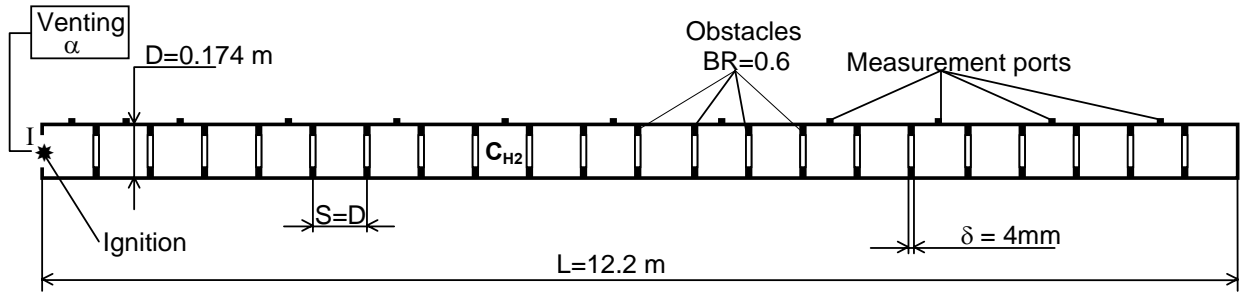

**Figure 1: Combustion tube configuration**

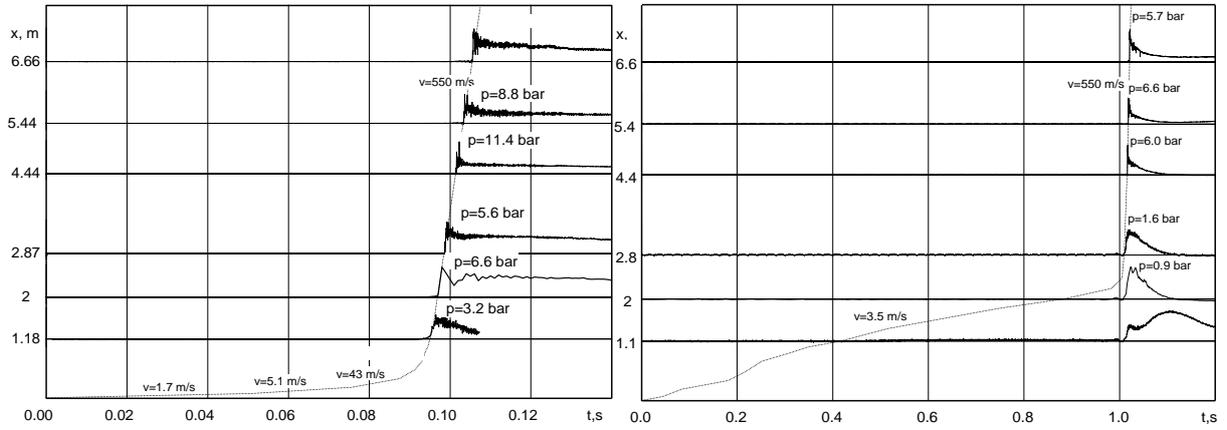

**Figure 2: x-t – diagram of the initial flame propagation in closed (left) and 40%-vented (right) obstructed channels: flame front (FF) trajectory (blue dots); light signals (black lines); pressure records (red lines). Pressure peaks and local visible flame velocities are shown at the plot.**

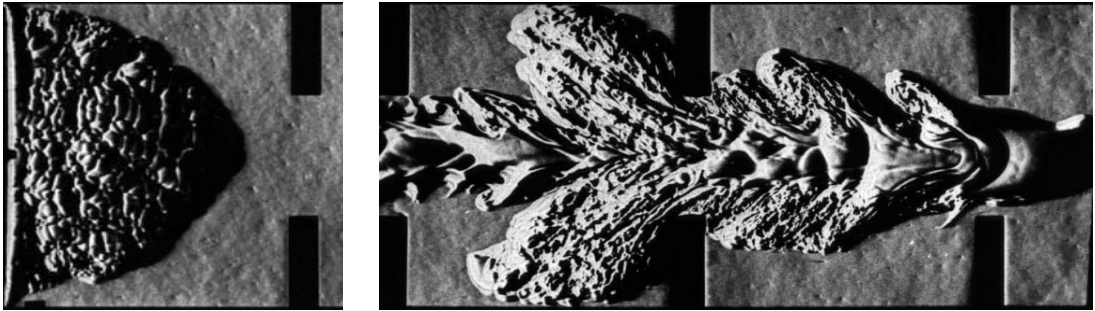

Figure 3: Laminar flame front structure for tested mixture (13% $H_2$-air, Le=0.36, closed tube, ignition at the end flange) in case of thermo-diffusive (left) and Kelvin-Helmholtz (right) instabilities.

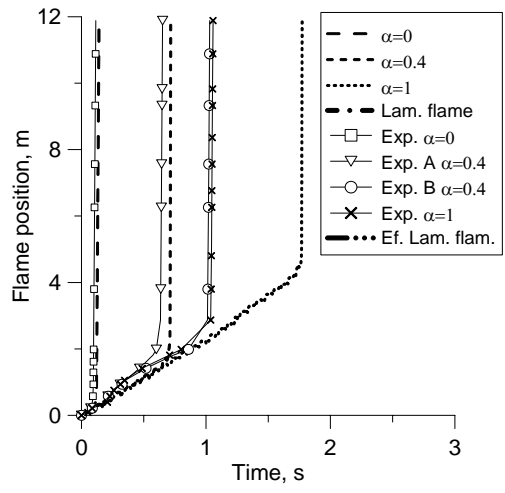

**Figure 4: Flame position. Dashed lines calculations. Thin continuous lines with symbols, experiments. Thick lines, laminar and quasi laminar regime propagation. Venting ratio is indicated in the legend.**

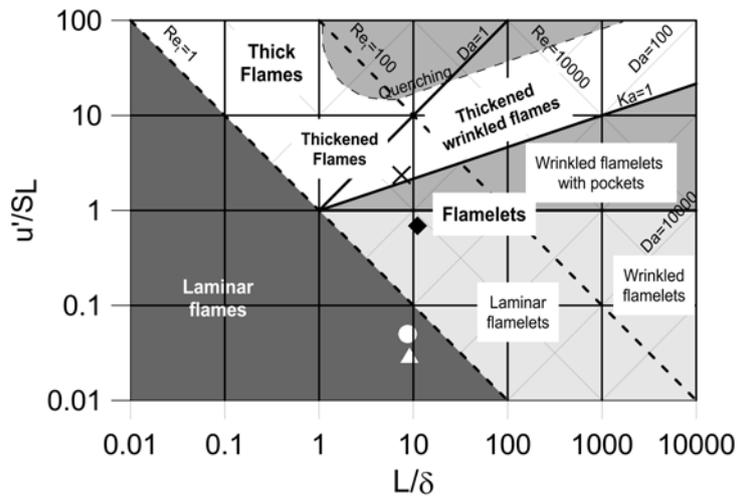

**Figure 5: Borghi Diagram. Diamond α=0, circle α=0.4, triangle α=1, cross *choked* regime.**

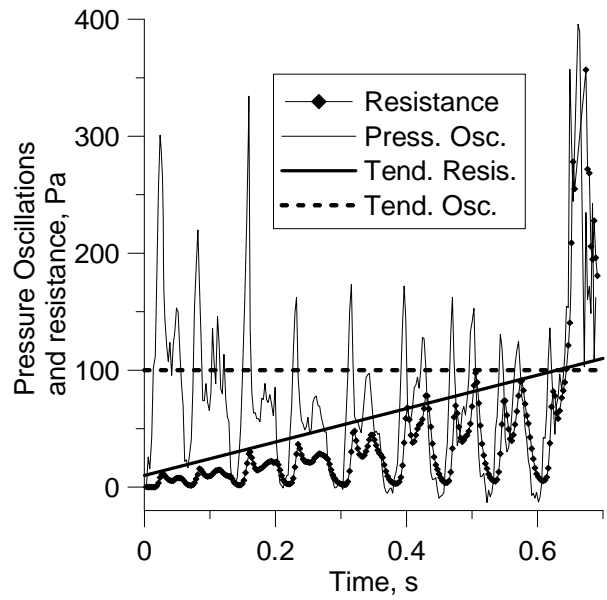

**Figure 6: Pressure oscillations and resistance obtained for the case α=s0.4.**

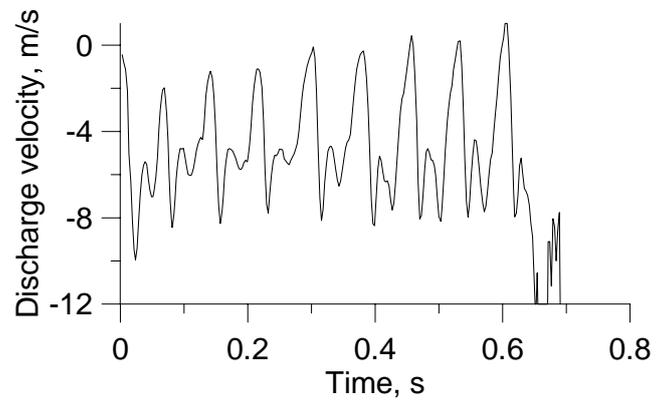

497
498 **Figure 7: Discharge velocity in venting orifice for the case α=0.4.**
499

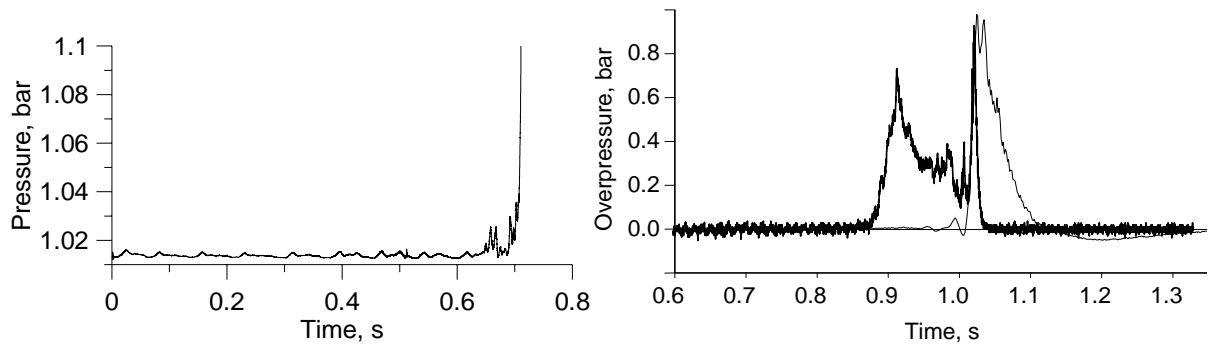

**Figure 8: Pressure records. Left, numerical simulation at the moment of the flame acceleration. Signal clipped at 1.1 bar. Right, experimental. Light signal (thin), pressure signal (thick).**

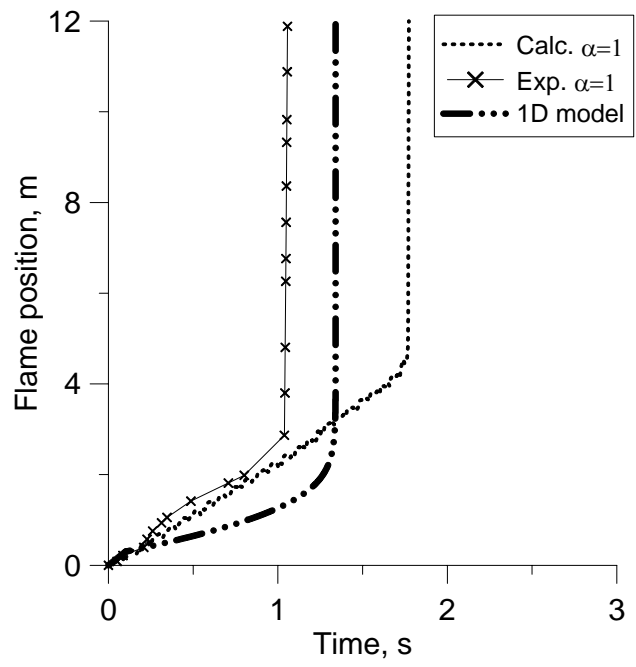

**Figure 9: Flame position. Comparison of calculations experiments and one dimensional model.**

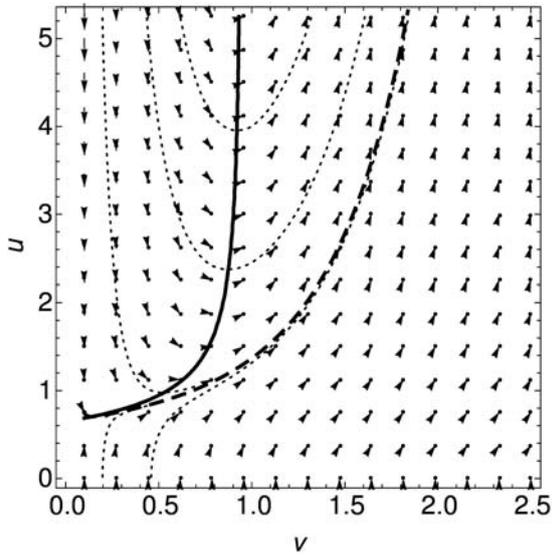 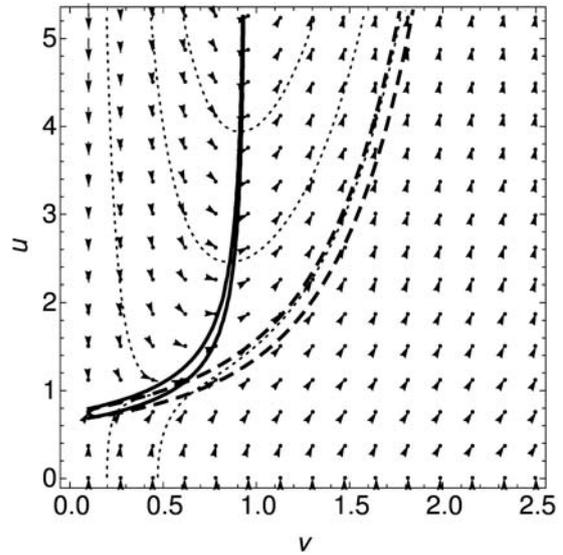

509

510 **Figure 10: Phase portrait $(x, \dot{x}) = (v, u)$ of the reduced model is shown with system solution trajectories**
511 **and a vector field. Green line shows the isoclines of the reaction wave speed showing minimal possible flame**
512 **velocity for a given initial state x. Red and blue lines are streamlines of the vector field. Dashed line is the**
513 **solution trajectory of the system. On the right both original and perturbed systems phase portraits are shown**
514 **for $\alpha = (1, 0.75)$ and $S_L^2 \to \alpha S_L^2$.**

515

**Table 1: Characteristics of the gases before flame arrival. Different degrees of venting and *choked* regime.**

| Venting | Ka | Da | Le | Ma | $S_L$ (m/s) | σ | Combustion regime |
|---|---|---|---|---|---|---|---|
| 0.0 | 1.92 | 15.9 | 0.36 | 1.13 | 0.23 | 3.38 | Laminar flamelets |
| 0.4 | 0.03 | 188 | 0.36 | 1.13 | 0.23 | 3.38 | Laminar flames |
| 1.0 | 0.01 | 327 | 0.36 | 1.13 | 0.23 | 3.38 | Laminar flames |
| - | 9.91 | 3.18 | 0.36 | 1.13 | 0.23 | 3.38 | Choked flame |